\documentclass[a4paper,12pt]{article}

\usepackage{jcappub}

\usepackage[british]{babel}
\usepackage[utf8]{inputenc}
\usepackage{amsfonts,amsmath,amssymb}
\usepackage{graphicx}
\usepackage{subfig}
\usepackage{hyperref}

\newcommand{\hebar}[0]{{}^3\overline{\text{He}}}
\newcommand{\tbar}[0]{\overline{\text{T}}}
\newcommand{\pbar}[0]{\overline{{p}}}
\newcommand{\nbar}[0]{\overline{{n}}}
\newcommand{\p}[0]{{p}}
\newcommand{\pbarp}[0]{\pbar/\p}

\newcommand{\qqbar}[0]{q\bar{q}}

\newcommand{\sigmav}[0]{\langle\sigma v\rangle}
\newcommand{\sol}[0]{\odot}
\newcommand{\Tn}[0]{T/A}

\newcommand{\defineText}[2]{\newcommand{#1}[0]{\text{#2}}}

\newcommand{\ccffqqqq}[0]{{\chi\chi\rightarrow\phi\phi\rightarrow\qqbar\qqbar}}
\newcommand{\ccqq}[0]{{\chi\chi\rightarrow\qqbar}}
\newcommand{\fqq}[0]{{\phi\rightarrow\qqbar}}

\defineText{\MV}{MV}
\defineText{\GV}{GV}
\defineText{\MeV}{MeV}
\defineText{\GeV}{GeV}
\defineText{\TeV}{TeV}
\newcommand{\GeVn}[0]{\GeV/\text{n}}

\defineText{\cm}{cm}
\defineText{\km}{km}
\defineText{\kpc}{kpc}

\defineText{\sr}{sr}

\defineText{\LIS}{LIS}
\defineText{\TOA}{TOA}

\newcommand{\pythia}[0]{{\texttt{PYTHIA}}}

\newcommand{\figurewidth}[0]{0.7\textwidth}

\title{Enhanced cosmic-ray antihelium production from dark matter annihilation through light mediators}

\author[a,b]{Yu-Chen Ding}   \emailAdd{dingyuchen@itp.ac.cn}
\author[c]{Nan Li}           \emailAdd{linan90@ihep.ac.cn}
\author[a,d,e]{Yu-Feng Zhou} \emailAdd{yfzhou@itp.ac.cn}

\affiliation[a]{CAS Key Laboratory of Theoretical Physics, Institute of Theoretical Physics, Chinese Academy of Sciences, Beijing 100190, China}
\affiliation[b]{School of Physical Sciences, University of Chinese Academy of Sciences, Beijing 100049, China}
\affiliation[c]{Key Laboratory of Particle Acceleration Physics and Technology, Institute of High Energy Physics, Chinese Academy of Sciences, Beijing 100049, China}
\affiliation[d]{School of Fundamental Physics and Mathematical Sciences, Hangzhou Institute for Advanced Study, UCAS, Hangzhou 310024, China}
\affiliation[e]{International Centre for Theoretical Physics Asia-Pacific, Beijing/Hangzhou, China}

\abstract{
Cosmic-ray (CR) antihelium is an important probe for the indirect search of dark matter (DM) annihilation in the Galaxy.
However, due to stringent constraints  from the measurements of CR antiprotons and $\gamma$-rays,
the flux of CR antihelium  from the conventional DM direct annihilation into Standard Model final states is expected to be  far below the sensitivity of  the current experiments. 
We show that the production of antihelium  can be significantly enhanced if the DM particles annihilate through  light mediator particles with a mass $m_\phi\approx8~\GeV$ close to the antihelium production threshold.
After taking into account the  constraints from the AMS-02 antiproton data and the Fermi-LAT $\gamma$-ray data on the spheroidal dwarf galaxies, we find that  in this scenario the CR antihelium flux  can be enhanced by three orders of magnitude, which  makes it within the sensitivity of the ongoing AMS-02 experiment.
}

\begin{document}
\maketitle
\flushbottom

\section{Introduction}
\label{sec:intro}

Astrophysical and cosmological observations suggest that $\sim 85\%$ of the matter in the present-day Universe is made of dark matter (DM). However, the particle nature of DM remains largely unknown. For decades, great efforts have been made to understand the particle nature of DM through direct, indirect, and collider detection experiments. 
If DM particles in the Galactic halo can annihilate (or decay) into the Standard Model (SM) final states, they can make extra contributions to the flux of cosmic-ray (CR) particles, which can be probed by high-precision DM indirect search experiments.

CR antiparticles 
are expected to be relatively rare as they are dominated by  CR secondaries produced  from the collisions between  primary CRs and the interstellar gas. Thus the CR antiparticle flux should be sensitive to the  contributions from DM interactions.
In recent years, a number of experiments  have detected an 
unexpected rise in the CR positron flux with kinetic energy above $\sim$10~GeV%
~\cite{Beatty:2004cy,PAMELA:2008gwm,Fermi-LAT:2011baq,AMS:2014bun}.
Halo DM annihilation  has been considered as a possible explanation 
(see e.g. Refs.~\cite{Kopp:2013eka,Bergstrom:2013jra,Ibarra:2013zia,Jin:2013nta}
for discussions related to the recent AMS-02 data) which is, however, 
subject to stringent constraints from the observations of $\gamma$-rays from dwarf galaxies (dSphs)~\cite{Fermi-LAT:2016uux}, the Galactic center~\cite{HESS:2016mib}, and the measurement of the anisotropy in the cosmological microwave background (CMB)~\cite{Planck:2015fie}.
A consistent DM explanation including the DM thermal relic abundance  may require complicated temperature dependence of DM annihilation, such as that with  $p$-wave Sommerfeld and resonant enhancements~\cite{Ding:2021zzg,Ding:2021sbj}.
%
%
Another important observable is the CR antiproton which is the lightest CR antinucleus and 
has been measured by a number of experiments such as 
PAMELA~\cite{Adriani:2012paa}, BESS-polar II~\cite{Abe:2011nx} and AMS-02~\cite{AMS:2016oqu}.  
The high precision data of AMS-02 show that in a large rigidity range from $\sim 1$ to 450~GV, 
the CR antiproton flux is in overall agreement with the secondary origin of CR antiprotons, 
which can be used to place stringent constraints on the possible hadronic interactions of DM particles
(see e.g. Refs.%
~\cite{
	Giesen:2015ufa,%
	Jin:2015sqa,%
	Lin:2016ezz,%
	Reinert:2017aga}).

Despite very small production rates, heavier antinuclei such as antideuteron 
($\overline{\textrm{D}}$) and antihelium-3 ($^3\overline{\textrm{He}}$) can also be 
useful probes of DM interactions.
This is because the  CR secondary antinuclei are suppressed 
in the low kinetic energy region  below $\sim\!\text{GeV/n}$ 
as the  antinuclei produced by $pp$-collisions are always highly boosted 
due to  the  high production  thresholds (17$m_{p}$ for  $\overline{\textrm{D}}$ and 
31$m_{p}$ for $^{3}\overline{\text{He}}$, where $m_{p}$ is the proton mass). 
The extremely low background makes it possible to search for the DM contributions in the low-energy region. 
Furthermore, at very high energies the secondary production is suppressed again by  the 
rapid falling of the primary CR flux at high energies (the CR proton flux scales 
with energy $E$ as $E^{-2.75}$), which opens another possible window for DM searches
with energy above $\sim 100~\text{GeV/n}$~\cite{Ding:2020puh}.

However, for benchmark DM  models where halo DM particles annihilate {\it directly} into SM quark pairs with a typical cross section of $\langle \sigma v\rangle \sim \mathcal{O}(10^{-26})~\text{cm}^3 \text{s}^{-1}$, the predicted $^{3}\overline{\text{He}}$ flux is known to be quite low \cite{Carlson:2014ssa,Cirelli:2014qia}.
Since in a given DM annihilation model, the production rates of CR antiheliums and CR antiprotons are strongly correlated, 
the upper limits on the CR antihelium flux can be estimated from the measured CR antiproton flux, 
which is highly insensitive to the DM annihilation cross section and the details of the CR propagation process~\cite{Ding:2018wyi}. 
It was shown  that for DM direct annihilation into quark pairs, 
the maximally allowed CR antihelium flux is below the sensitivity of the AMS-02 experiments by 
roughly two orders of magnitude~\cite{Ding:2018wyi}.

Recently, the AMS-02 collaboration has shown 
several preliminary CR antihelium candidate events
at rigidities below 50~GV~\cite{samuelting2016}, 
which has motivated theoretical efforts to look for 
mechanisms that can increase the antihelium production rate from either DM or exotic sources~\cite{Coogan:2017pwt,Poulin:2018wzu,Cholis:2020twh,Winkler:2020ltd,Kachelriess:2021vrh}. 
The recently proposed approaches for increasing the antihelium production rate typically involve nonstandard values of some phenomenological parameters. 
For instance, a relatively large coalescence momentum of $p^{A=3}\approx 0.36~\text{GeV}$ for 
 $^{3}\overline{\text{He}}$ formation was adopted in~\cite{Coogan:2017pwt},
while the values extracted  from the ALICE data is $p^{A=3}\approx 0.25~\text{GeV}$.
In Ref.~\cite{Cholis:2020twh}, a
large  Alfvén wave velocity of $\sim 60~\text{km/s}$, which controls the reacceleration of CR particles during propagation~\cite{Cholis:2020twh} was considered. However, global fits to the CR data typically give $20-40~\text{km}\cdot\text{s}^{-1}$~\cite{Strong:1998pw,Jin:2014ica,Cuoco:2019kuu}.
In Ref.~\cite{Winkler:2020ltd} a large value of di-quark production rate which controls the hadronization process in the $\pythia$ package was adopted, corresponding to the parameter value of $\texttt{probQQtoQ} \sim 0.24$.
A large di-quark production rate can enhance the production of the $\bar{\Lambda}_b$ baryon which can subsequently decay into $^{3}\overline{\text{He}}$.
Note, however, that the default value of {\tt probQQtoQ} in $\pythia$ was only 0.09 as referenced in Ref.~\cite{Winkler:2021cmt}, which has been further reduced to 0.081 since version {\tt8.204} to agree with the Monash 2013 tune~\cite{Skands:2014pea}.
It was shown that  increasing the  di-quark production rate will affect all the predicted baryon and meson production rates, which is inconsistent with the experiment data. For instance, the corresponding proton multiplicity predicted at $\sqrt{s}=91~\text{GeV}$ is about $33~\sigma$ away from the measurement~\cite{Kachelriess:2021vrh}.   

In this work, we show that CR antihelium produced by DM annihilation can be naturally enhanced if the DM particles annihilate into SM quarks through a mediator particle with a mass $m_\phi\approx8~\GeV$, which is slightly above the $\hebar$ production threshold $E^{(\hebar)}_{\text{thr}} \approx 5.6~\GeV$.
We find that compared with the direct DM annihilation scenario, 
the DM antihelium flux can be enhanced up to three orders of magnitude 
under the constraints from the AMS-02 antiproton-to-proton ratio data~\cite{AMS:2021nhj} and the Fermi-LAT dSph $\gamma$-ray observations~\cite{Fermi-LAT:2013sme,Fermi-LAT:2015att}.
We show that the AMS-02 experiment is capable of detecting the enhanced DM antihelium flux in this scenario. 
We also show that  the mass ratio between the mediator and DM particles can be inferred from the observed energy spectrum of the CR antihelium events.

This paper is structured as follows.
In \autoref{sec:prod}, we review the coalescence model for CR antihelium production, and then calculate the energy spectra of antiprotons and antiheliums from DM annihilation using the Monte-Carlo simulation package $\pythia$ and the coalescence model on an event-by-event basis.
In \autoref{sec:galprop},
the CR propagation model used to predict the fluxes of antiprotons and antiheliums is described.
In \autoref{sec:limits}, we calculate the upper limits on the cross section of DM annihilation through light mediator particles using the AMS-02 $\pbarp$ data and Fermi-LAT $\gamma$-ray data from dSph galaxies.
In \autoref{sec:prospect}, we discuss the prospect of detecting antiheliums from DM annihilation at the AMS-02 experiment.
The conclusion is given in \autoref{sec:conclusions}.

\section{Enhanced antihelium production from DM annihilation through light mediator}
\label{sec:prod}

In this work, we investigate the $\hebar$ production in the process where two non-relativistic DM particles with mass $m_\chi$ annihilate into a pair of neutral mediator particles with mass $m_\phi$,
which subsequently decay into SM particles through the $\qqbar$ channel, namely $\ccffqqqq$.
For simplicity, we only consider the light quarks $q=u,d,s$.
Extending the analysis to other heavy flavors is straightforward,
and the results are expected to be similar.
The production of CR $\hebar$ from DM annihilation can be described as the coalescence of antiprotons ($\pbar$) and antineutrons ($\nbar$) produced by the annihilation.
They can be formed from the coalescence of two antiprotons and one antineutron, or two antineutrons and one antiproton as antitritium nuclei ($\tbar$) which decay subsequently into $\hebar$.

\begin{figure}[t]
    \centering
    \includegraphics[width=\figurewidth]{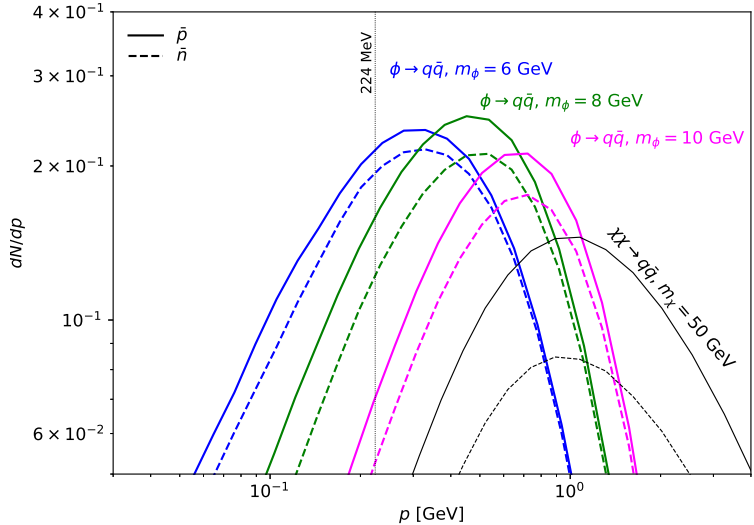}
    \caption{
    Distribution of $\pbar$ and $\nbar$ produced by the mediator decay (colored lines) and the direct DM annihilation (black lines) processes with respect to the nucleon momentum $p$ in the center-of-mass frame of $\qqbar$, simulated with $\pythia$.
    The vertical line indicates the typical coalescence momentum $p_0^{\hebar} = 224~\MeV$ for $\hebar$ production.
    }
    \label{fig:npbars-production}
\end{figure}

In the center-of-mass frame of the $\qqbar$ system, the kinetic energies of the produced antinucleons are correlated to the mediator particle mass $m_\phi$.
In general, one expects a softer antinucleon spectrum from a lighter mediator particle.
We calculate the energy spectra of antinucleons from DM annihilation using the Monte Carlo method by utilizing the $\pythia$ event generator of version {\tt8.306}.
The mediator $\phi$ is treated as a non-resonance particle,
which is the default setup in $\pythia$ for $m_\phi < 20~\GeV$.
The simulated spectra of antinucleons produced by the decay of light mediator particles with $m_\phi=6-10~\GeV$ are compared with that by the direct annihilation of DM with $m_\chi=50~\GeV$ in \autoref{fig:npbars-production}.
The figure shows that,
although the light mediator decay processes with less center-of-mass energy tend to produce fewer antinucleons than that in the direct annihilation scenario in total numbers,
they produce more antinucleons at low momentum region of $p\lesssim\mathcal{O}(10^{-1})~\GeV$,
which is of particular importance when considering antinucleus formation.

A coalescence model
\cite{Butler:1963pp,Schwarzschild:1963zz,Csernai:1986qf}
is adopted to simulate the formation of $\hebar$ during the DM annihilation process.
In this model, all possible combinations of $\pbar\,\pbar\,\nbar$ and $\pbar\,\nbar\,\nbar$ from one DM annihilation event are investigated using coalescence criteria concerning their relative momentum and spatial configuration.

The momentum criterion is introduced geometrically.
We first calculated the three invariant relative momenta,
$k_1 = \sqrt{(p_2 - p_3)^2}$, $k_2 = \sqrt{(p_3 - p_1)^2}$ and $k_3 = \sqrt{(p_1 - p_2)^2}$,
for each combination of triple antinucleons with four momenta $p_1$, $p_2$ and $p_3$.
Then a triangle is constructed with the lengths of its sides taking the value of the three invariant relative momenta,
and is checked if it can be enveloped by a circle with diameter parameter $p_0^{\bar{A}}$,
where $\bar{A}$ is $\hebar$ for $\pbar\,\pbar\,\nbar$ and $\tbar$ for $\pbar\,\nbar\,\nbar$.
In cases where the triangle of relative momenta is obtuse or can not be constructed,
i.e., $k_i^2 \ge k_j^2 + k_l^2$ with the longest edge as $k_i$,
the criterion is trivially
\begin{equation}
    k_i \le p_0^{\bar{A}}, \quad \text{for} ~ i = 1, 2, 3
    \label{eq:envelop1}.
\end{equation}
In other cases where the triangle is acute,
a weaker inequality is used where $p_0^{\bar{A}}$ is compared with the diameter of the triangle's circumcircle, i.e.,
\begin{equation}
    d_\text{circ} =
    \frac{k_1 k_2 k_3} {\sqrt{ (k_1+k_2+k_3) (k_1+k_2-k_3) (k_2+k_3-k_1) (k_3+k_1-k_2) }} \le p_0^{\bar{A}}
    \label{eq:envelop2}.
\end{equation}

In addition to the momentum criterion,
we also require the coalescing antinucleons have relatively close spatial distances.
The spatial criterion is implemented by grouping the final state antinucleons according to the lifetime of their parent particles.
The decay of an intermediate state particle is treated as a transient process if its lifetime is shorter than a lifetime cutoff $\tau_0$,
while the decay products of a long-lived particle with a lifetime longer than $\tau_0$ are considered off-vertex,
meaning that the produced antinucleons are too far away to form antinuclei.
Note that, it is possible that a long-lived intermediate state particle that decays off-vertex can produce enough antinucleons to form antinuclei by itself.
Such secondary contributions are particularly important for processes with heavy initial states through heavy flavor channels, such as the $\bar{\Lambda}_b$ from $b\bar{b}$ channel~\cite{Winkler:2020ltd}.
In this work, the off-vertex produced antinuclei are taken into account in the production spectra,
even though the actual contribution is negligible for the $\qqbar$ channels.

If all the considered criteria are met, we count one antinucleus produced from the tested group of antinucleons.
The four-momentum of the produced antinucleus is identified as
\begin{equation}
    p^{\bar{A}} = m_A \frac{ p_1+p_2+p_3 } {\sqrt{( p_1+p_2+p_3 )^2}},
\end{equation}
which is according to the center-of-mass velocity of the antinucleus system.

In the coalescence model, the overall multiplicity of antinuclei produced by DM annihilation is affected by two competing factors, the number of possible antinucleon combinations, and the coalescence probability. From the simulation results shown in \autoref{fig:npbars-production},
one can see that heavier mediator particles can decay into more antinucleons,
leading to more possible combinations of antinucleons for coalescence.
However, the produced antinucleons from heavier mediator particles tend to have larger relative momenta, which suppresses the possibility of passing the momentum criterion in the coalescence model.

The total $\hebar$ production spectrum is the sum of directly coalesced $\hebar$ nuclei and those from the decay of produced $\tbar$ nuclei.
We take
$\tau_0 = 2 \times 10^{-12} ~\text{mm}/c$,
and the center value of $p_0^{\hebar} = 224^{+12}_{-16}~\MeV$ and $p_0^{\tbar} = 234^{+17}_{-29}~\MeV$
determined from our previous study~\cite{Ding:2018wyi} using the $pp$-collision data from the ALICE experiment~\cite{ALICE:2017xrp}.
In \autoref{fig:npbars-production}, we compare $p_0^{\hebar}$ with the momentum distributions of antinucleons produced by mediator decay and DM annihilation.
It can be roughly viewed as a typical momentum upper bound for antinuclei allowed to coalesce into antiheliums.

The detailed statistics of the simulated $\hebar$ produced by the decay of mediator particles ($\fqq$) and the direct annihilation of DM particles with mass $m_\chi=50~\GeV$ are listed in \autoref{tab:sim-detail}.
The table shows that decays of light mediator particles can produce more $\hebar$ than the direct annihilation process up to $2-3$ orders of magnitude.
There exists a mediator mass $m_\phi\sim8~\GeV$ that can yield the highest amount of $\hebar$ during their decay.
At this mass value, a balance is reached between the number of possible antinucleon combinations and the coalescence probability.

\begin{table}[t]
\centering
\begin{tabular}{cccc}
\hline\hline
$E_{\text{cm}}$ [GeV]
& \begin{tabular}[c]{@{}c@{}}Simulated\\Events\end{tabular}
& \begin{tabular}[c]{@{}c@{}}Total Number\\of $\hebar$\end{tabular}
& Multiplicity \\
\hline
($\fqq$, $E_{\text{cm}}=m_\phi$) \\
$ 6$  & $10^{10}$  & 18226           & $1.82\times10^{-6}$ \\
$ 7$  & $10^{10}$  & 110680          & $1.11\times10^{-5}$ \\
$ 8$  & $10^{10}$  & 145732          & $1.46\times10^{-5}$ \\
$ 9$  & $10^{10}$  & 63524           & $6.35\times10^{-6}$ \\
$10$  & $10^{10}$  & 11717           & $1.17\times10^{-6}$ \\
\hline
($\ccqq$, $E_{\text{cm}}=2m_\chi$) \\
$100$ & $10^{11}$ & 4746            & $4.75\times10^{-8}$ \\
\hline
\end{tabular}
\caption{
Simulated numbers of $\hebar$ production from light mediator decay ($\fqq$) and direct DM annihilation ($\ccqq$) at different center-of-mass energies $E_{\text{cm}}$, using the $\pythia$ event generator and the coalescence model.
}
\label{tab:sim-detail}
\end{table}

\begin{figure}[t]
    \centering
    \includegraphics[width=\figurewidth]{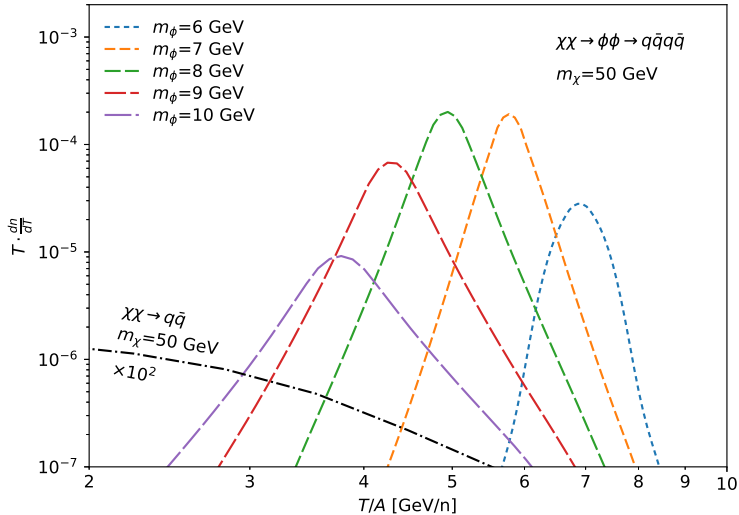}
    \caption{
    $\hebar$ spectra produced by the annihilation of two 50 GeV DM particles through light mediator particles (colored dashed lines) compared with the direct annihilation production (black dash-dotted lines), simulated with $\pythia$.
    Contributions from $\tbar$ decay are included.
    The $\hebar$ spectrum from direct DM annihilation (black dash-dotted line) is scaled up by $10^2$ for comparison.
    }
    \label{fig:hetbars-production}
\end{figure}

For DM annihilation processes,
the existence of light mediator particles also leads to a distinct feature in the $\hebar$ energy spectra.
In the rest frame of $\phi$,
the produced $\hebar$ tends to have low kinetic energies as $T^{(\phi)}_{\hebar} \lesssim m_\phi - 2 m_{\hebar}$.
While in the center-of-mass frame of DM annihilation,
the $\hebar$ nuclei are boosted by the Lorentz factor of the light mediator $\gamma_\phi = m_\chi / m_\phi$,
resulting in a highly concentrated distribution at a typical total energy
\begin{equation}
E_{\text{typical}} \approx m_{\hebar} \gamma_\phi = m_{\hebar} m_{\chi} / m_{\phi}.
\end{equation}
In \autoref{fig:hetbars-production}, we show the production spectra of $\hebar$ from the annihilation of two $50~\GeV$ DM particles.
The kinetic energy per nucleon corresponding to the typical total energy of the DM produced $\hebar$ can be recognized from the peak structure of the distribution.

\section{Propagation of CR antinuclei}
\label{sec:galprop}

The propagation of charged cosmic rays in the Galaxy can be described by the diffuse-reacceleration model \cite{1994ApJ...431..705S,Strong:1998pw,Moskalenko:2001ya,Strong:2007nh,Maurin:2001sj,Maurin:2002ua}.
The diffusion zone is modeled as a cylinder with a radius $r_h = 20~\kpc$ and half-height $z_h = 1\sim10 ~\kpc$.
In the diffusion zone, the evolution equation for the phase space distribution $f(\vec{r}, p)$ of CR particles can be written as
\begin{equation}
    \frac{\partial f}{\partial t}
    = q(\vec{r}, p)
    + \nabla \cdot(D_{xx} \nabla f - \vec{V}_c f)
    + \frac{\partial}{\partial p} p^2 D_{pp} \frac{\partial}{\partial p} \frac{f}{p^2}
    - \frac{\partial}{\partial p} \left[ \dot{p} f - \frac{p}{3}(\nabla\cdot\vec{V}_c)f\right]
    - \frac{f}{\tau_f}
    - \frac{f}{\tau_r}
    \label{eq:diff-reacc},
\end{equation}
where
$D_{xx}$ is the spatial diffusion coefficient that depends on the rigidity $R=p/|Z|$ of the CR particle as
$D_{xx} = \beta D_0(R/R_0)^\delta$,
where $D_0$ is the normalization factor,
$\delta$ is the spectral power index,
and $\beta=v/c$ is the velocity of CR particles in the unit of speed of light.
$D_{pp} = 4V_a^2 p^2/(3D_{xx}\delta(4-\delta^2)(4-\delta))$ is the diffusion coefficient in momentum space with $V_a$ as the Alfvén velocity.
$\vec V_c$ is the convection velocity driven by the Galactic wind.
$\dot p$ represents the energy loss rate of CR nuclei during propagation, including the ionization and Coulomb interaction with interstellar medium (ISM) particles~\cite{Strong:1998pw}.
$\tau_f$ and $\tau_r$ are the average lifetimes for fragmentation and decay respectively.
In this work, the reacceleration effect, controlled by the parameter $V_a$, is considered  a constant in the diffusion zone.
We noticed that in some studies the reacceleration is confined in a thin disk of half-height $h_r$, and $V_a$ takes a different value.
It has been shown that the effect of different treatments of the reacceleration region can be effectively absorbed into $V_a$, as $V_a^2\propto h_r$~\cite{Maurin:2002hw}.

The CR source term $q(\vec r, p)$ represents the injected number of CR particles per unit volume, momentum, and time.
For primary CR nuclei, it is the product of
a broken power-law spectrum as a function of particle rigidity,
and a spatial function that follows the distribution of supernova remnant \cite{1996A&AS..120C.437C}, i.e.,
\begin{equation}
    q_\text{nuc} (\vec r, p) \propto
    \left( \frac{R}{R_\text{br}} \right)^\nu
    \left( \frac{r}{r_\sol} \right)^{1.25}
    \exp \left( - 3.56 \frac{r - r_\sol}{r_\sol} \right)
    \exp \left(-\frac{|z|}{0.2~\kpc} \right)
    \label{eq:qnuc},
\end{equation}
where $r_\sol = 8.5~\kpc$ is the Galactocentric distance of the sun, and $\nu = \nu_1, \nu_2$ for rigidity $R$ below and above the breaking rigidity $R_\text{br}$ respectively.

The CR source term for $\pbar$ and $\hebar$ produced by the annihilation of Majorana DM particles with mass $m_\chi$ and velocity averaged cross section $\sigmav$ is given by
\begin{equation}
    q (\vec r, p) = \frac {\rho^2_\text{DM}(\vec r)} {2 m_\chi^2} \sigmav \frac{dN}{dp}.
\end{equation}
We take the Navarro-Frenk-White (NFW) profile \cite{Navarro:1996gj} for the DM density distribution which is parameterized as
\begin{equation}
    \rho(r)_\text{DM} = \rho_\odot \frac{r_\odot}{r}
    \left( \frac {r_\odot + r_s} {r + r_s} \right)^2,
\end{equation}
with $r_s = 20~\kpc$ and the local DM density $\rho_\odot = 0.43~\GeV/\cm^3$.
The momentum distributions ${dN}/{dp} = \beta ({dN}/{dT})$ are calculated using the event-by-event Monte-Carlo method described in \autoref{sec:prod},
where the kinetic energy is defined in the rest frame of the DM halo.
For DM annihilation with mediator particles, the $\hebar$ spectra from mediator decay are boosted by the Lorentz factor of the mediator particle $\gamma_\phi = m_\chi / m_\phi$, and multiplied by a factor of two as two mediator particles decayed in a single DM annihilation.

The secondary $\pbar$ and $\hebar$ are produced as primary CR nuclei scattering with ISM.
The secondary source term for the antiparticle $\bar{A}$ can be written as
\begin{equation}
    q_{\bar{A}}
    (\vec r, p) =
    \sum_{ij} n_j(\vec r)
    \int dp'_{i} ~
    n_i(\vec r, p'_i) ~
    \beta_i ~
    \sigma_\text{inel}^{ij}(p'_i)
    \frac{dN_{\bar{A}}}{dp},
\end{equation}
where the indices for CR species $i$ and for ISM component $j$ are summed through ${}^1\text{H}$ and ${}^4\text{He}$ nuclei.
In this work, we adopt the secondary production spectrum ${N_{\hebar}}/{dp}$
calculated with the MC event generator \texttt{EPOS-LHC} \cite{Pierog:2013ria} as presented in our previous work \cite{Ding:2018wyi}.
For the secondary antiproton source term, we adopt the parameterization of $\pbar$ production cross section implemented in the $\texttt{GALPROP}$ code,
which is based on Ref.~\cite{Tan:1983kgh,Barashenkov:1994cp}.

The fragmentation rates for CR antinuclei with mass number $A$ are proportional to their inelastic cross section with ISM hydrogen and helium nuclei, which can be written as
\begin{equation}
    \frac{1}{\tau_f} = (n_\text{H}
    + \frac{\sigma_{\bar{A}\alpha}}{\sigma_{\bar{A}p}} ~ n_\text{He})
    ~ v ~ \sigma_{\bar{A}p}
    \label{eq:frag-rate}.
\end{equation}
Here we take the geometrical factor $\sigma_{\hebar\alpha}/\sigma_{\hebar p} = 1.98$
as empirically determined in Ref.~\cite{Ferrando:1988tw}.
Assuming CP-invariance,
the inelastic cross sections of $\bar{A}p$-collision are parameterized as~\cite{MOISEEV1997379,Moskalenko:2001ya}
\begin{equation}
\sigma_{\bar{A}p} = \sigma_{\bar{p}A} =
A^{2/3} [48.2 + 19 x^{-0.55} + (0.1-0.18 x^{-1.2})|Z| + 0.0012 x^{-1.5}Z^2] ~\text{mb}
\label{eq:sigmaAp},
\end{equation}
where $Z$ is the atomic number of the nucleus,
and $x$ is the kinetic energy per nucleon of the projectile in units of $\GeVn$.
The inelastic cross section for antiprotons interacting with ISM protons is taken from Ref.~\cite{Tan:1983de,Moskalenko:2001ya}.

We solve Eq.~\eqref{eq:diff-reacc} numerically using the code \texttt{GALPROPv54}~\cite{Strong:2007nh}.
The parameters for the primary nuclear sources and CR propagation are taken
as that obtained by fitting the CR proton flux and B/C ratio to the AMS-02 data \cite{Jin:2014ica},
which are listed in \autoref{tab:params}.
The local interstellar fluxes with respect to kinetic energy per nucleon $\Tn$ are extracted at $\vec r = \vec r_\odot$ as
$\Phi^{\text{LIS}}(\Tn) = \frac{Ac}{4\pi} f(\vec r_\odot, p)$.

For the modulation effect on CR fluxes from the magnetic field in the solar system,
$\Phi^\text{LIS}$ is transformed into the
fluxes at the top of Earth's atmosphere $\Phi^\text{TOA}$ using the force-field approximation
\begin{equation}
    \Phi_{A,Z}^\TOA (T^\TOA) =
    \left(\frac{p^2_\TOA}{p^2_\LIS}\right) \Phi_{A,Z}^\LIS (T^\LIS)
    \label{eq:force-field},
\end{equation}
where a CR particle with atomic number $Z$
loses energy by $T^\LIS - T^\TOA = \phi_F e {|Z|}$ as it propagates to Earth.
In this work, the value of $\phi_F$ is fixed at $550~\MV$~\cite{Jin:2014ica}.

\begin{table}[t]
\centering
\begin{tabular}{ccccccccc}
\hline\hline
  $r_h~[\kpc]$ &
  $z_h~[\kpc]$ &
  $R_0~[\GV]$ &
  $D_0~[\cm^2/\text{s}]$ &
  $\delta$ &
  $V_a~[\km/\text{s}]$&
  $R_\text{br}~[\GV]$ &
  $\nu_1$ &
  $\nu_2$ \\
  \hline
$20$ &
  $3.2$ &
  $4$ &
  $6.5\times10^{28}$ &
  $0.29$ &
  $44.8$ &
  $10$ &
  $1.79$ &
  $2.45$
\\\hline
\end{tabular}
\caption{
CR propagation parameters and primary nucleus source parameters used in this work.
These are the benchmark values obtained by fitting the AMS-02 B/C ratio and proton data based on the \texttt{GALPROPv54} code \cite{Jin:2014ica}.
}
\label{tab:params}
\end{table}

\section{Constraints from CR antiproton and $\gamma$-ray data}
\label{sec:limits}

For DM annihilation through hadrophilic mediators, the DM annihilation cross section is subject to stringent constraints from the data of CR antiprotons and $\gamma$-rays.
In this section, we derive the constraints from the AMS-02 antiproton-to-proton ($\pbarp$) ratio data~\cite{AMS:2021nhj} and the Fermi-LAT dSph $\gamma$-ray observation~\cite{Fermi-LAT:2013sme,Fermi-LAT:2015att}.

In this work, the constraints on DM annihilation cross section from CR $\pbarp$ and dSph $\gamma$-rays are derived individually,
and we do not attempt to combine them.
The reason is that the prediction of these two observables suffers from different sources of uncertainties.
The uncertainty of the predicted CR $\pbarp$ ratio lies in the CR propagation model,
which is strongly correlated with the prediction of CR $\hebar$ flux.
While for dSph $\gamma$-ray, the uncertainty mainly comes from the DM profile of dSphs.

\subsection{Constraints from AMS-02 antiproton data}
\label{sec:limit-pbarp}

Since the production of CR $\hebar$ from DM annihilation strongly correlates with CR antiproton creation, it is stringently constrained by experimental observations of CR antiprotons.
In this work, we use the seven-year data of the $\pbarp$ flux ratio from the AMS-02 experiment to constrain the DM annihilation cross sections.
Following the method described in \autoref{sec:galprop},
we solve the propagation equations for the secondary and DM produced $\pbar$ along with primary protons.
The ratio of antiproton and proton fluxes predicted by the numerical simulation is then compared with the AMS-02 $\pbarp$ data.

A frequentist $\chi^2$-analysis is adopted to derive the upper limits of the DM annihilation cross section.
The $\chi^2$ function is defined as $\chi^2 = \sum_i(\Phi_i^{\text{th}} - \Phi_i^{\text{exp}})^2/\sigma_i^2$,
where the index $i$ goes through each energy bin of the AMS-02 $\pbarp$ data, $\Phi_i^{\text{th}}$ and $\Phi_i^{\text{exp}}$ are the model predicted and AMS-02 observed CR antiproton fluxes in each energy bin respectively, and $\sigma_i$ are the experimental uncertainties.
For fixed values of DM and mediator masses,
we first find the value of $\sigmav$ that minimize the $\chi^2$ function to $\chi^2_\text{min}$.
Then we determine the upper limit of DM annihilation cross section at 95\% confidence level (C.L.) by finding the $\sigmav$ that gives $\Delta\chi^2 = \chi^2 - \chi^2_\text{min} = 2.71$ corresponding to a single degree of freedom.

In our analysis, we have fixed the propagation parameters and DM halo profile to the model described in \autoref{sec:galprop}.
The predictions of the maximal CR $\hebar$ flux produced by DM annihilation are highly insensitive to these models, due to the fact that a variation in these models mainly leads to a rescaling of the DM annihilation cross sections in such a way that the same antiproton flux is reproduced~\cite{Ding:2018wyi}.

\begin{figure}[t]
    \centering
    \includegraphics[width=\figurewidth]{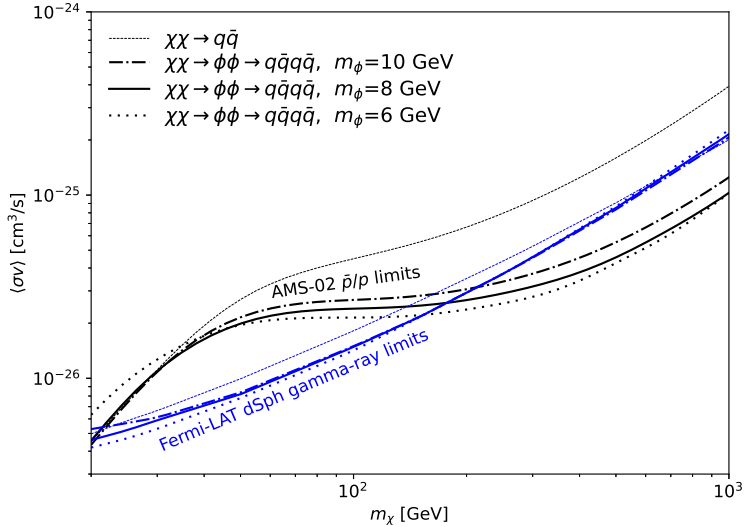}
    \caption{
    95\% C.L. upper limits on DM annihilation cross section $\sigmav$ from the data of AMS-02 $\pbarp$ ratio (black lines) and Fermi-LAT dSphs $\gamma$-rays (blue lines) as functions of DM mass $m_\chi$ for different annihilation scenarios.
    }
    \label{fig:limits}
\end{figure}

In \autoref{fig:limits} we show the 95\% C.L. upper limits of the DM annihilation cross section through $\qqbar$ channels constrained by AMS-02 $\pbarp$ data.
As shown in the figure, for $m_\chi\lesssim100~\GeV$, the upper limits for DM annihilating through light mediators are comparable with those without mediators, but have slightly stronger constraints for DM mass above $\sim100~\GeV$.

\subsection{Constraints from Fermi-LAT dSphs $\gamma$-ray observations}
\label{sec:limit-dsphs}

The Fermi-LAT collaboration published the $\gamma$-ray energy fluxes of 25 Milky Way dSphs~\cite{Fermi-LAT:2015att}.
The fluxes were calculated from the PASS8 analysis using the six years of Fermi-LAT data,
with $\gamma$-ray energy ranging from 500 MeV to 500 GeV.
In this work, we choose the 15 dSphs that have independent sky regions and $J$-factor uncertainties \cite{Fermi-LAT:2013sme} to place constraints on the DM annihilation cross sections.

The predicted energy flux of $\gamma$-rays produced by Majorana DM annihilation from the halo of the $i$th dSphs between the $j$th energy bin $(E_{\min,j}, E_{\max,j})$ can be written as
\begin{equation}
\phi_{i,j} (m_\chi, \langle \sigma v \rangle, J_i) =
\frac{J_{i}}{4\pi}
\frac{ \langle \sigma v \rangle }{ 2 m_{\chi}^{2} }
\int _{E_{\min,j}} ^{E_{\max,j}} dE_{\gamma}
\ E_{\gamma} \frac{dN}{dE_{\gamma}}_{\chi\chi\rightarrow\gamma}.
\label{eq:dsphs-eflux}
\end{equation}
The factor $J_i$ represents the geometric information of the DM mass density profile $\rho_i(\vec{r})$ of the dSphs, which is defined by a line-of-sight integral inside the region-of-interest $\Delta \Omega_i$.
For annihilating DM particles,
\begin{equation}
    J_i = \int_{\Delta \Omega_i} d \Omega \int_0^\infty dl \rho_i^2(\vec{r}).
\end{equation}
The mean values of $\overline{\log J_i}$ and their uncertainties $\Delta \log J_i$ for the 15 dSphs are listed in \autoref{tab:dsphs}.

In Eq.~\eqref{eq:dsphs-eflux}, we have assumed that the considered dSph halos share the same $\sigmav$ as the Milky Way halo.
It is worth noting that in some DM models, the annihilation cross section becomes velocity dependent due to the Sommerfeld and resonant enhancements~\cite{Ding:2021zzg,Ding:2021sbj}, and must be treated individually for each DM halo.
This effect is not significant in this work, as the Sommerfeld effect saturates to be velocity independent when the typical DM velocity $v_0/c \ll m_\phi/m_\chi$.

\begin{table}[t]
\centering
\begin{tabular}{lcc}
\hline\hline
Name              & $\overline{\log_{10}J}$ & Ref.                   \\ \hline
Bootes I          & $18.8 \pm 0.22$         & \cite{DallOra:2006tki} \\
Canes Venatici II & $17.9 \pm 0.25$         & \cite{Simon:2007dq}    \\
Carina            & $18.1 \pm 0.23$         & \cite{Walker:2008ax}   \\
Coma Berenices    & $19.0 \pm 0.25$         & \cite{Simon:2007dq}    \\
Draco             & $18.8 \pm 0.16$         & \cite{Munoz:2005be}    \\
Fornax            & $18.2 \pm 0.21$         & \cite{Walker:2008ax}   \\
Hercules          & $18.1 \pm 0.25$         & \cite{Simon:2007dq}    \\
Leo II            & $17.6 \pm 0.18$         & \cite{Koch:2007ye}     \\
Leo IV            & $17.9 \pm 0.28$         & \cite{Simon:2007dq}    \\
Sculptor          & $18.6 \pm 0.18$         & \cite{Walker:2008ax}   \\
Segue 1           & $19.5 \pm 0.29$         & \cite{Simon:2010ek}    \\
Sextans           & $18.4 \pm 0.27$         & \cite{Walker:2008ax}   \\
Ursa Major II     & $19.3 \pm 0.28$         & \cite{Simon:2007dq}    \\
Ursa Minor        & $18.8 \pm 0.19$         & \cite{Munoz:2005be}    \\
Willman 1         & $19.1 \pm 0.31$         & \cite{Willman:2010gy}  \\ \hline
\end{tabular}
\caption{
Names and $J$-factors (in units of $\log_{10}[\GeV^2\cm^{-5}]$) of the 15 dSphs used to constrain the DM annihilation cross section.
The mean values and uncertainties of $\log_{10} J$ are taken from Ref.~\cite{Fermi-LAT:2013sme}.
}
\label{tab:dsphs}
\end{table}

The overall likelihood function for the observed $\gamma$-ray energy fluxes from all dSphs can be written as
\begin{equation}
\mathcal{L} (m_\chi, \langle \sigma v \rangle) \propto
\prod_{i}^{N_\text{dSphs}}
\prod_{j}^{N_\text{bins}}
\mathcal{L}_{i,j}(\phi_{i,j}(m_\chi,\langle \sigma v \rangle, \hat{J}_{i}))
\times
e^{-\frac12\left(\frac{\log \hat{J}_i - \overline{\log J_i}}{\Delta \log J_i}\right)^2}
\label{eq:dsphs-like}
\end{equation}
where the likelihood function for the $i$th dSphs in $j$th energy bin, $\mathcal{L}_{i,j}$,
as a function of the predicted energy flux $\phi_{i,j}$ are obtained using the pre-calculated log-likelihood table \cite{Fermi-LAT:2015att}.
$\hat{J}_i$ are the values of $J$-factors that maximize the likelihood function for each dSph.

To determine the upper limits of the averaged DM annihilation cross sections at 95\% C.L.,
we define the test statistics
\begin{equation}
    \operatorname {TS} = - 2 \ln \frac
    { \mathcal L (m_\chi, \sigmav) }
    { \mathcal L (m_\chi, \sigmav_0) }
    \label{eq:dsphs-ts},
\end{equation}
with $\sigmav_0$ the averaged DM annihilation cross section that maximize $\mathcal L(m_\chi, \sigmav)$.
For each $m_\chi$ in each annihilation scenario,
we find the values of $\sigmav_0$ that maximize $\mathcal{L}(m_\chi, \sigmav_0)$, and the 95\% C.L. upper limits of $\sigmav$ is found by searching up from $\sigmav_0$ for a value that yields $\operatorname{TS}=2.71$.

In \autoref{fig:limits} we compare the 95\% C.L. upper limits for DM annihilation cross sections of different annihilation scenarios constrained independently by the AMS-02 $\pbarp$ ratio and Fermi-LAT $\gamma$-ray observations on the 15 dSphs.
In the direct annihilation scenario $\ccqq$, the dSph $\gamma$-rays place more stringent constraints than the $\pbarp$ ratio for all DM masses above 20 GeV.
In the light mediator scenario,
the constraints from $\pbarp$ are stronger than that from dSphs $\gamma$-rays for $m_\chi \gtrsim 200~\GeV$.
While for $20~\GeV \lesssim m_\chi \lesssim 200~\GeV$, it is still dSph $\gamma$-rays data that place stronger constraints on the DM annihilation cross sections.
Note that some previous studies have suggested an  excess in the AMS-02 $\pbarp$ data at $\sim10~\GeV$, which can be possibly explained by a DM annihilation signal~\cite{Cui:2016ppb,Cuoco:2016eej,Reinert:2017aga,Cuoco:2017okh}.
However, other analyses showed that the significance of such an excess is quite low after consistently considering the uncertainties from the $\pbar$ production cross section, CR propagation model, and possible correlation errors in the AMS-02 data~\cite{Calore:2022stf,Luque:2021ddh,Heisig:2021ujl,Heisig:2020nse,Cuoco:2019kuu}.
\autoref{fig:limits} shows that at $\sim 10$~GeV the constraints from $\bar p$ data are relatively weak, but more stringent constraint may arise from the dSph $\gamma$-ray data, which  disfavors the DM interpretation for the $\pbarp$ spectrum at a few GeV.
We leave a fully combined analysis including both $\bar p$ and $\gamma$-ray data for future work.

\section{Prospect of detecting antiheliums from DM annihilation through light mediators}
\label{sec:prospect}

\begin{figure}[t]
    \centering
    \includegraphics[width=\figurewidth]{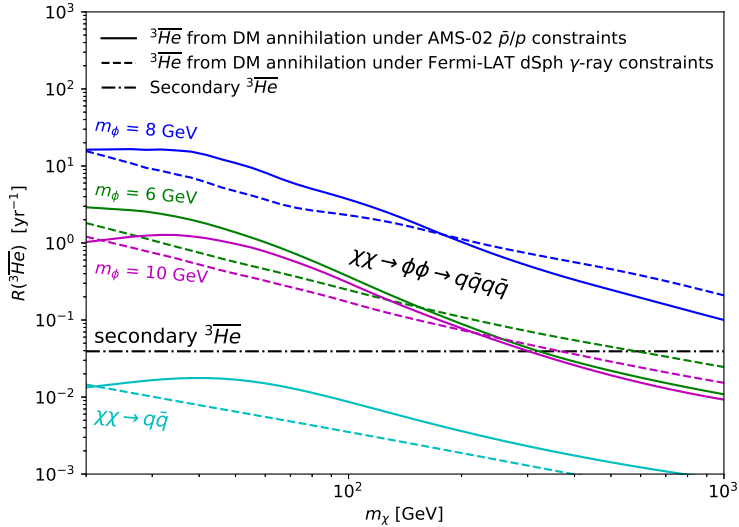}
    \caption{
    Maximal CR $\hebar$ event rates as functions of DM mass $m_\chi$ at AMS-02 for different annihilation scenarios (distinguished by colors).
    The DM annihilation cross sections are taken to be the 95\% C.L. upper limits given by AMS-02 $\pbarp$ flux ratio (in solid lines) and Fermi-LAT dSphs $\gamma$-rays (in dashed lines).
    The horizontal dash-dotted line indicates the contribution from the secondary $\hebar$ produced by $pp$-collisions during CR propagation.
    }
    \label{fig:ams-rate}
\end{figure}

After obtaining the constraints on the DM annihilation cross section in \autoref{sec:limits}, we estimate the event rates of DM induced $\hebar$ particles that can be detected by the AMS-02 experiment.
The observed $\hebar$ event rate $\mathcal{R}$ is estimated as the integral of CR $\hebar$ flux multiplied by the acceptance $\mathcal{A}$ and efficiency $\eta$ of the detector, i.e.,
\begin{equation}
    \mathcal{R} = \int d T ~ \eta ~ \mathcal{A} ~ \Phi_{\hebar}.
    \label{eq:event-rate}
\end{equation}
As an optimistic estimation, we take the geometric acceptance of the AMS-02 detector $\mathcal{A} = 0.5~\text{m}^2\cdot\text{sr}$~\cite{AMS:2002yni} and $\eta = 1$ in the range of nucleon kinetic energy from $0.1~\GeVn$ to $1~\TeV/\text{n}$ during 18 years of AMS-02 operation.
We show the maximally expected $\hebar$ event rates on AMS-02 as a function of DM mass in \autoref{fig:ams-rate}.
Compared to that from direct DM annihilation, the event rate of $\hebar$ can be enhanced up to $2\sim3$ orders of magnitude if DM particles annihilate through mediator particles with $m_\phi \lesssim 10~\GeV$, and can be above the secondary $\hebar$ backgrounds.

\begin{figure}[t]
    \centering
    \includegraphics[width=\figurewidth]{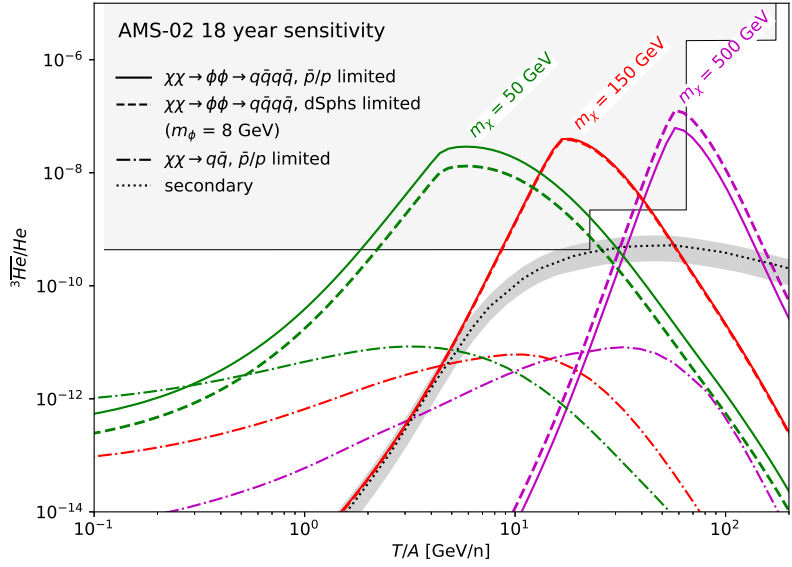}
    
    \caption{
    Flux ratios of $\hebar$ from DM annihilation through $8~\GeV$ mediator particles over $\text{He}$ compared with the AMS-02 18 years sensitivity, with the DM annihilation cross sections constrained by
    the AMS-02 $\pbarp$ ratio data (solid lines) and
    the Fermi-LAT dSphs $\gamma$-ray observations (dashed lines).
    Three different DM masses $m_\chi=50~\GeV$, $150~\GeV$, and $500~\GeV$ are distinguished in color.
    For comparison, the flux ratios with direct annihilating DM constrained by AMS-02 $\pbarp$ data are shown in dot-dashed lines.
    The ratio for secondary $\hebar$ produced by primary CR particles colliding with the ISM is shown in the dotted line with a gray band indicating the uncertainty from the coalescence model~\cite{Ding:2018wyi}.
    The shaded area indicates the AMS-02 18 years sensitivity for $\hebar$/$\text{He}$ ratio \cite{Kounine:2010js}.
    }
    \label{fig:hebar-ratio}
\end{figure}

In \autoref{fig:hebar-ratio}, the predicted flux ratios of $\overline{{}^3\text{He}}/\text{He}$ are compared with 
a more realistic estimation of the AMS-02 detecting sensitivity during its 18-year operation
\cite{Kounine:2010js}.
As shown in the figure, without involving mediator particles, $\hebar$ signals from DM direct annihilation are about two orders of magnitude below the detecting sensitivity to be distinguished from CR $\text{He}$.
Secondary $\hebar$ can marginally reach the detecting sensitivity at kinetic energy per nucleon $\Tn\gtrsim 10~\GeV$.
The AMS-02 experiment after 18 years of operation is capable of detecting $\hebar$ from DM annihilation through light mediators with $m_\chi \lesssim 200~\GeV$ and $m_\phi = 6\sim10~\GeV$ if the DM annihilation cross section takes the 95\% C.L. upper limits allowed by the current results of $\pbarp$ ratio and dSphs $\gamma$-ray observations.
In the latter scenario, AMS-02 is expected to detect a peak in the $\overline{\text{He}}/\text{He}$ ratio.
This feature corresponds to the boost effect of the light mediator particles (as discussed in \autoref{sec:prod} and shown in \autoref{fig:hetbars-production}).
One can estimate the typical energy of the detected $\hebar$ using the mass ratio of DM and mediator particles as
$E_\text{typical} \approx m_{\hebar} m_\chi / m_\phi$.

\section{Conclusions}
\label{sec:conclusions}

In summary, we have shown that DM annihilation through light mediator particles can produce detectable cosmic-ray $\hebar$ particles.
A light mediator can significantly reduce the relative momenta of produced antinucleons and hence increase the possibility of antinucleus formation.
By adopting an event-by-event Monte Carlo method with \texttt{PYTHIA8} and the coalescence model,
we calculated the expected CR $\hebar$ flux from DM annihilation channel $\chi\chi\rightarrow\qqbar$ and $\chi\chi\rightarrow\phi\phi\rightarrow\qqbar\qqbar$
under the limits from the AMS-02 antiproton and Fermi-LAT dSphs $\gamma$-ray observation.
We found that, under both constraints,
with DM particles of mass $m_\chi \lesssim 10^2~\GeV$ annihilating through light mediator particles of mass $m_\phi \approx 8~\GeV$,
the DM annihilation through light mediator particles can produce detectable CR $\hebar$ for the AMS-02 experiment.
Another difference between DM annihilation with and without mediator particles is the energy distribution of the produced CR $\hebar$ flux.
While in the direct annihilation scenario DM annihilation produces $\hebar$ with a relatively flat spectrum,
the CR $\hebar$ particles from DM annihilation through light mediator are expected to have a typical total energy
$E_\text{typical} \approx m_{\hebar} m_\chi / m_\phi$.
Hence, it is possible to extract
the mass ratio of the mediator and DM particles $m_\phi/m_\chi$
from the observed energy distribution of CR $\hebar$.

\section*{Acknowledgements}
This work is supported in part by
the National Key R\&D Program of China No.~2017YFA0402204,
the CAS Project for Young Scientists in Basic Research YSBR-006,
the National Natural Science Foundation of China (NSFC)
No.~11825506,  
No.~11821505,  
and                                            
No.~12047503.  

\bibliographystyle{utcaps}
\bibliography{refs}

\end{document}